\documentclass[nofootinbib,a4paper,aps,amsmath,amsfonts,twocolumn]{revtex4}
\usepackage[mathscr]{eucal}
\def\Pcm#1{{\mathcal{#1}}}
\newcommand{\del}{\partial}
\def\eqref#1{(\ref{#1})}
\def\er#1{eqn.\eqref{#1}}
\def\nn{\nonumber}
\begin{document}
\title{Our response to the response {\texttt{hep-th/0608109}} by Drummond}
\author{N.~D.~Hari Dass} 
\email{hari@soken.ac.jp}
\affiliation{Hayama Center for Advanced Studies, Hayama, Kanagawa, Japan}
\author{Peter Matlock} 
\affiliation{Department of Electrophysics, National Chiao Tung University, Hsinchu, Taiwan}
\begin{abstract}
We have carefully examined all the points raised by Drummond in his
response \cite{drumresp} to our paper \cite{orig} wherein we had made
some criticisms of his earlier work \cite{drum}. We concede that
Drummond is correct in claiming the non-existence of $R^{-4}$- and $R^{-5}$-order 
effective string actions in the parity conserving sector, though
only insofar as equivalence of field theories is considered at the
classical level; the situation in unclear when quantum equivalence is
taken into consideration.  We still maintain the existence of such
terms in the parity violating sector. Nevertheless we point out that
all this has no consequence for our original proof of the nonexistence
of order-$R^{-3}$ terms.  Apart from this we refute Drummond's claims
about our alleged use of field redefinitions as well as his criticism
of our dropping $R^{-4}$ terms in our analysis. We reject his
contention that our work is merely a partial reconstruction of his
original results and that our work contains technical and conceptual
errors.  We do acknowledge the importance of the absence of terms
pointed out by Drummond.
\end{abstract}
\maketitle


\section{Preliminary comments}
\label{intro}
There are several components to Drummond's response to our work. We
appreciate the very professional nature of his response as well as the
fact that he has given full details of all his derivations making it
easy for us to be focussed in our `response to the response'. We now
give our replies, but before doing so point out a few typos in our
original paper.
\subsection{Typographical errors in our original paper}
\label{errors}
Eqn(17) of \cite{orig} should read:
\begin{eqnarray}
\label{R3EOM}
&&\frac{2}{a^2}\del_{+-}Y^\mu =
- 4\frac{\beta}{R^2}\del_+^2\del_-^2 Y^\mu \nn\\
&&\quad-4\frac{\beta}{R^3}\bigg[
\del_+^2\big\{
              \del_-^2Y^\mu(e_+\cdot\del_-Y+e_-\cdot\del_+Y)
                \big\} \nn\\
        &&\quad\quad+\del_-^2\big\{
               \del_+^2Y^\mu(e_+\cdot\del_-Y+e_-\cdot\del_+Y)
                 \big\}  \nn\\
&&\quad\quad -  e_+^\mu\del_-(\del_+^2\cdot\del_-^2Y)-e_-^\mu\del_+(\del_+^2Y\cdot\del_-^2Y)
\bigg]
.\end{eqnarray}

Eqn(18) of \cite{orig} should read:
\begin{eqnarray}
\frac{2}{a^2}\del_{-}Y_1^\mu &=& 4\frac{\beta}{R^3}
\big( e_+^\mu\del_+Y_0\cdot\del_-^3Y_0+e_-^\mu\del_+^2Y_0\cdot\del_-^2Y_0 \nn\\
&-&\del_-^2Y_0^\mu e_-\cdot\del_+^2Y_0-\del_+Y_0^\mu e_+\cdot\del_-^3Y_0 \big) 
.\end{eqnarray}

\subsection{Fluctuation Field Y and X-uniformity}
\label{fluctyxuni}
With no loss of generality one can make a functional shift and set
$X^\mu = X_{\textup{cl}}^\mu + Y^\mu$ where $X_{\textup{cl}}^\mu =
R(e_+^\mu\tau^++e_-^\mu\tau^-)$ which is a solution of the equation of
motion (EOM) of the free part of the action. Since all additional
terms in the effective action involve higher derivatives,
$X_{\textup{cl}}^\mu$ continues to be a solution of the full EOM, and
$Y^\mu$ continues to have the interpretation of a fluctuation around a
classical background.

While carrying out field redefinitions of the $Y$-field, care should
be taken not to upset the $(0,0)$ nature of the $X$-field. As this property
is not explicit in the $Y$-formulation, it is very easy to upset.

Since the fluctuation field $Y$ derives from the $X$-field, one could
consider the following principle of ``$X$-uniformity'': \emph{All
expressions involving $Y^\mu$ and its derivatives, $e_{\pm}^\mu, R$
must be such they are derivable from expressions involving $X^\mu$ and
its derivatives only}.

Many field redefinitions of the $Y$-field will not be permissible if
this principle is adopted. In fact the field redefinitions found by
Drummond involving $R^{-2}$ and $R^{-3}$ terms of the
Polchinski-Strominger (PS) action are of a type that do not satisfy
this principle.

If this principle of $X$-uniformity is applied, the $(0,0)$ property
of $Y^\mu$ is never upset. In fact, without this principle, it seems
unclear how to define precisely the $(0,0)$ property for the
fluctuation field. Still, this does not necessarily mean that such a
definition is impossible in case $X$-uniformity is not imposed.

Further motivation for the imposition of $X$-uniformity is as
follows. When it is broken, the fluctuation field $Y$ and derived
spectrum thereof no longer have a direct interpretation in terms of
excitations about some background, since of course it is not possible
in such a case to write down the background in question.  Without such
an interpretation, the physical significance or meaning of the
fluctuations may be difficult to appreciate.

On the other hand, if one is given an effective theory describing
fluctuations, there is \emph{a priori} no clear reason why one should
not define the field differently, and indeed one could set out
with such an effective theory without even having seen the original 
$X$ field. 

\section{Equivalence of field theories}
\label{equivft}
As many of the issues raised in this discussion centre around the
issue of equivalence of field theories under a change of variables
(what is called field redefinition) we review here known facts and
establish some new points in the context of effective string theories.

As far as effective string theories are concerned two distinct
possibilities exist as to their interpretation as quantum
theories. One is to treat them as a tool to calculate the so-called
tree diagrams only.  In the early days of effective chiral symmetric
field theories (of pions, for example) this was the attitude taken; in
this case, equivalence of field theories would only mean equivalence
as in classical field theories, as explained below in section
\ref{ecf}.

Now, it is well known in chiral perturbation theory that such a
limited approach would have missed many essential features like
universal logarthmic behaviour and would have made a discussion of
analytic properties of scattering amplitudes beyond perview. The more
modern approach is to make the calculation of even loop amplitudes
meaningful and the lack of renormalisability is handled through a
larger set of arbitrary parameters. If we interpret the effective
string theories this way, all features of a full quantum field theory
are to be considered.

\subsection{Classical field theories}
\label{ecf}
Equivalence in classical field theories is easily proved; see for
example \cite{kamefuchi} for an explicit demonstration for the case of
\emph{point transformations}, namely, cases where the redefinition does
not involve time derivatives. At a classical level this can be
extended also to more general redefinitions.  This classical
equivalance means that for such things as the currents and
energy-momentum tensor, either using the lagrangian formalism on the
original lagrangian and then transforming the fields to a new set
gives the same results as first transforming the fields and then
applying the formalism.  The point is that the change of variables and
the lagrangian formalism commute, and a field redefinition can thus be
done at will to simplify the procedure without further thought.

When the redefinition in question is an \emph{infinitesimal} one it is
possible, classically, to eliminate terms in the action that are
proportional to the leading order (in an appropriate sense) EOM.
Equivalently, it can be stated that two classical actions differing
from each other by EOM terms in this sense are physically equivalent.
Therefore the PS prescription \cite{PS} of dropping all EOM terms as
\emph{irrelevant} is certainly justified at a classical level.  The
important question is to what extent this procedure is also valid
quantum-mechanically.  We are able to make some progress on this issue
at least within the path integral formulation. We argue, somewhat
formally, that up to order $R^{-3}$ classical equivalence also implies
quantum equivalence, but beyond that things are uncertain.  This means
that our original proof of the absence of order-$R^{-3}$ corrections
to the spectrum is still valid and complete. It also means that the
claim of absence of order-$R^{-4}$ and -$R^{-5}$ corrections in the
action made by Drummond in \cite{drum} is certainly true classically
but its validity quantum-mechanically is not certain.

\subsection{Canonical formulation of quantum field theories}
\label{cqft}
Proofs of equivalence in the canonical, or operator, formalism of
quantum field theories are not completely straightforward, the main
obstacle being the non-commutativity of various operator expressions;
this has been very clearly analysed in \cite{kamefuchi}. Although in
that paper they do manage to prove the quantum equivalence for point
transformations, such a proof for field redefinitions that go beyond
point transformations does not seem to exist. In the context of
effective string theories the field redefinitions considered are
certainly not of the point-transformation type.

We shall therefore analyse this issue within the path integral
formalism. There is of course an intimate connection (and usually
assumed equivalence) between the operator and path integral
formulations of quantum field theories. Though there are very
important structural and conceptual differences, the final results for
physically relevant issues are expected to be the same. This
equivalence between canonical and path integral formulations is subtle
and delicate as shown long ago by Lee and Yang \cite{leeyang}.

\subsection{Path integral formulation}
\label{pif}
In the path integral formulation, given a particular field definition,
there are three important issues: the action; the transformation laws
leaving the action invariant; the invariance of the measure under said
transformation laws. For non-linear transformation laws the
tranformation of the measure can of course be quite involved.

In the case of the PS transformation law, it is easy to see that the
na\"ive measure ${\cal D}X$ is indeed not invariant. We calculate now
the change in the na\"ive measure under the PS transformation
law. Consider the relation between the untransformed field $Y^\mu$ and
the (infinitesimally) transformed field ${Y^\prime}^\mu$;
\begin{widetext}
\begin{eqnarray}
{Y^\prime}^\mu &=& Y^\mu+R\epsilon^-~e_-^\mu+\epsilon^-\partial_-~Y^\mu+\frac{\beta~a^2}{R}e_+^\mu\partial_-^2\epsilon
+\frac{\beta~a^2}{R^2}\partial_-^2\epsilon^-\{\partial_+Y^\mu+2e_+^\mu(e_-\cdot\partial_+Y+e_+\cdot\partial_-Y)\}\nonumber\\
&+&\frac{2\beta~a^2}{R^3}\partial_-^2\epsilon^-e_+^\mu\{2(e_+\cdot\partial_-Y+e_-\cdot\partial_+Y)^2+\partial_+Y\cdot\partial_-Y\}
+\cdots
.\end{eqnarray}
Denoting the na\"ive measure by ${\cal D}Y$, its change under the above transformation is
${\cal D}Y^\prime = {\cal J}{\cal D}Y$,
where ${\cal J}$ is the determinant of 
${\cal J}^\mu_\nu(\xi,\xi^\prime)= \frac{\delta {Y^\prime}^\mu(\xi^\prime)}{\delta Y^\nu(\xi)}$
given by
\begin{eqnarray}
{\cal J}^\mu_\nu(\xi,\xi^\prime) &=&\delta^\mu_\nu\delta^2(\xi-\xi^\prime)+\delta^\mu_\nu\epsilon^-\partial_-\delta^2(\xi-\xi^\prime)
+\frac{\beta~a^2}{R^2}\delta^\mu_\nu\partial_-^2 \epsilon^-\partial_+\delta^2(\xi-\xi^\prime)+\frac{2\beta~a^2}{R^2}\partial_-^2\epsilon^-
(e_+^\mu e+^\nu\partial_-\delta^2+e_+^\mu e_-^\nu\partial_+\delta^2)\nonumber\\
&+&\frac{2\beta~a^2}{R^3}\partial_-^2\epsilon^-(\delta_\nu^\mu\partial_+\delta^2
+4e_+^\mu(e_+^\nu\partial_-\delta^2+e_-^\nu\partial_+\delta^2))(e_-\cdot\partial_+Y+e_+\cdot\partial_-Y)
+e_+^\mu(\partial_-Y^\nu\partial_+\delta^2+\partial_+Y^\nu\partial_-\delta^2)\nonumber\\
&+& \frac{2\beta~a^2}{R^3}\partial_-^2\epsilon^-\partial_+Y^\mu(e_+\nu\partial_-\delta^2+e_-^\nu\partial_+\delta^2)+\cdots
.\end{eqnarray}
\end{widetext}
The appearance of $\delta^2(\xi-\xi^\prime)$ and its derivatives makes
the evaluation of ${\cal J}$ very delicate and a proper regularisation
scheme, like for example the $\zeta$-function regularisation, is
needed to do this carefully.  There are two possible fates for these
singular terms.

One possibility is that the singular terms from the measure could
cancel against singular non-covariant expressions arising on a careful
use of Feynman rules in the apparently covariant path integral. This
is what was shown in \cite{leeyang}. The same situation
shows up in perturbative quantum gravity also \cite{gdep}.

The other possibility is that there are singular terms in the measure
which do not vanish or cancel in the above fashion, leaving
non-trivial factors to be taken into account. The only relevant
changes are contained in the {\em $Y$-dependent} part of ${\cal J}$
and these come from those parts of the transformation law having at
least quadratic terms in the fluctuation field $Y^{\mu}$.

We conclude that at least formally, by which we mean that of course a
careful regularised definition of the path integral in principle
should be taken into account, one can say that the variation of the
measure under the PS transformation law is order-$R^{-3}$ and hence
does not affect our original work where only variations that are of
order $R^{-2}$ are relevant.  Nevertheless, the important point
evident from the above expression is that when any analysis is carried
out for order-$R^{-4}$ terms and beyond, this could become an
important issue and must be handled with care.

\subsection{Jacobian for field redefinitions}
\label{fredefj}
When a generic field redefinition (change of variables in the path
integral) is made, the new action is obtained from the old by simple
substitution. The new transformation laws are what are induced by the
redefinition. Unless the field redefinition is invertible it is not
possible to work out the induced transformation laws. For infintesimal
changes in the fields, this can always be done.

Of course, this is not the whole picture; the change in the measure
for path integration consequent to the change in variables should also
be taken into account.  Simple counting arguments of the type
developed by us in \cite{orig} and improved upon by Drummond in
\cite{drumresp} show that the most general field redefinitions have
the structure
\begin{equation}
\Delta Y^\mu = c_1(R) [Y] + c_2(R) [YY] + c_3(R) [YYY]+\cdots
\end{equation}
where $[Y],[YY],\dots$ symbolically denote terms that are linear,
quadratic, etc.~in the $Y$-field. Further it can be shown that the
coefficients $c_n(R)$ can {\em at most} be of order $R^{-(n+1)}$. The
dominant field redefinition in fact has the form
\begin{eqnarray}
{Y^\mu}^\prime = Y^\mu&+& \frac{\alpha}{R^2}\partial_{+-} Y^\mu  
+\frac{\beta}{R^3}\partial_{+-}Y^\mu (e_+\cdot\partial_-Y)
+\cdots \nn\\
&+&\frac{\gamma}{R^4}\partial_{+-}Y^\mu(\partial_+Y\cdot\partial_-Y)
+\cdots
\end{eqnarray}
The jacobian for this transformation can again be formally computed as
before. Again we encounter singular expressions and careful
regularisation is required to make further progress. The non-trivial
part of the jacobian, by which we mean the $Y$-dependent terms, can
only come from terms that are {\em cubic} in the $Y$-field in the
field redefinition and those are of order $R^{-4}$. Once again these
are irrelevant for the proof presented in \cite{orig}.

We conclude this section with two points.  Firstly, as far as the
proof of the absence of $R^{-3}$ corrections to the Nambu-Goto
spectrum is concerned, quantum equivalence is the same as classical
equivalence.

Secondly, if a careful evaluation of these jacobians leads to local
terms, one may be able to use Drummond's results for absence of terms
to order $R^{-5}$ to show that classical equivalence implies quantum
equivalence up to this higher order. A rough argument in support of
this could be constructed by making field redefinitions and performing
partial integration as done by Drummond to eliminate all terms up to
order $R^{-6}$. An inspection of the field redefinitions carried out
by Drummond in \cite{drumresp} reveals that at least formally the
resultant jacobian also does not get any contributions to order
$R^{-5}$. Nevertheless, this is only a sketch and we can have
confidence in it only after it is carefully established.  Until then,
quantum equivalence to orders $R^{-4}$ and beyond should be taken as
unproven.

\section{Order-$R^{-4},R^{-5}$ terms in the effective string action}
\label{rm4rm5}
In our proof of the absence of $R^{-3}$ corrections to the spectrum it
was necessary to prove the absence of any extra order $R^{-3}$ terms
in the action for effective strings beyond those already there by
carrying out the expansion of the Polchinski-Strominger action to this
order.  For this we developed a systematic way of analysing the order
at which various action terms could arise. In this analysis, in which
we considered both parity-conserving as well as parity-violating terms
(for the sake of completeness), we proved the result, crucial for
analysis, that there are no order $R^{-3}$ terms. This enabled us to
consistently expand the PS action \cite{PS} to order $R^{-3}$, and
their stress tensor to order $R^{-2}$.

Though we did not state it explicitly in \cite{orig}, this also
implies that it is enough to expand the transformation law given by PS
to include order-$R^{-2}$ terms. Incidentally, the transformation law
is closed (satisfying the Virasoro algebra) to order $R^{-3}$; only at
order $R^{-4}$ do the PS transformations fail to close.

Our analysis showed the potential existence of order $R^{-4},R^{-5}$
terms. In \cite{drum} Drummond had claimed, without any systematic
analysis, that the next corrections to the PS action would be order
$R^{-6}$ and he listed four such terms. We had criticised this in
\cite{orig}.

Now we have thoroughly re-examined this issue and agree with Drummond
that our potential $R^{-5}$ and $R^{-4}$ terms in the parity
conserving sector can all be reduced to order $R^{-6}$ terms modulo
irrelevant terms (which according to us can simply be dropped provided
the transformation laws are modified to keep all the relevant terms
invariant).

\subsection{Parity violating sector}
\label{pvs}
In the parity violating sector, our claims of the existence of
$R^{-4},R^{-5}$ terms remain valid. We deal with this issue in some
detail now. In the $D=3$ case we had shown that
\begin{equation}
\label{3dpv}
L^{-3}\epsilon_{\mu_1\mu_2\mu_3}\partial_+X^{\mu_1}\partial_-X^{\mu_2}\partial_+^2 X^{\mu_3}\partial_+X\cdot\partial_-^3 X
\end{equation}
is potentially of order $R^{-3}$. This had been omitted in
\cite{drum}. However around the classical background $X_{\textup{cl}}^\mu =
e_+^\mu R \tau^+ + e_-^\mu R \tau^-$,$X^{\mu}=X_{\textup{cl}}^\mu+Y^\mu$ this
becomes
\begin{equation}
-8R^{-3}\epsilon_{\mu_1\mu_2\mu_3}e_+^{\mu_1}e_-^{\mu_2}\partial_+^2 Y^{\mu_3}e_+\cdot\partial_-^3 Y
\end{equation}
which we had shown can be reduced by partial integration to 
\begin{equation}
8R^{-3}\epsilon_{\mu_1\mu_2\mu_3}e_+^{\mu_1}e_-^{\mu_2}\partial_+ Y^{\mu_3}e_+\cdot\partial_+\partial_-^3 Y
\end{equation}
and hence rendered irrelevant. We consequently dropped them (for
a discussion of Drummond's criticism of this and our defense, see
sections \ref{afr} and \ref{rmf}). As we were only interested in showing the
non-existence of {\bf relevant} $R^{-3}$ terms in \cite{orig}, this
was sufficient. In view of the present discussion we now analyse
the $R^{-4}$ and $R^{-5}$ terms also. For that we give an improved
treatment; by partial integration we can recast \eqref{3dpv} as
\begin{eqnarray}
&-&L^{-3}\epsilon_{\mu_1\mu_2\mu_3}\partial_-X^{\mu_2}\partial_-({\partial_+X^{\mu_1}\partial_+^2 
X^{\mu_3}})\partial_+ X\cdot\partial_-^2 X \nn\\
&-&L^{-3}\epsilon_{\mu_1\mu_2\mu_3}\partial_+ X^{\mu_1}\partial_-^2 X^{\mu_2}\partial_+^2 
X^{\mu_3}~\partial_+ X\cdot\partial_-^2 X  \\
&-&3L^{-4}\epsilon_{\mu_1\mu_2\mu_3}\partial_+ 
X^{\mu_1}\partial_- X^{\mu_2}\partial_+^2 X^{\mu_3}(\partial_-^2 X\cdot\partial_+X)^2 \nn
.\end{eqnarray}
The first line produces irrelevant terms of order $R^{-3}$ and
higher. We will discuss later how these and other irrelevant terms are
treated while responding to criticisms on the use of field
redefinitions. Both the remaining lines are of order $R^{-4}$ and
higher. It should be noted that this has been done without any
redefinition of fluctuation field and thus automatically obeys our
principle of $X$-uniformity introduced in section \ref{fluctyxuni}.

If we do make a redefinition of the $Y$ field, we may further recast the $R^{-4}$ terms as
\begin{equation}
\frac{8}{R^4}\epsilon_{\mu_1\mu_2\mu_3}e_+^{\mu_1}\partial_+^2Y^{\mu_3}e_+\cdot\partial_-^2Y\big\{\partial_-^2 Y^{\mu_2}
 - 6 e_-^{\mu_2}e_+\cdot\partial_-^2 Y\big\}
\end{equation}
These can again be reduced to irrelevant terms by partial integration,
and therefore $R^{-4}$ terms may also be eliminated, at the expense 
of $X$-uniformity.
Nevertheless, the $R^{-5}$ terms remain and we see no way to get rid
of them, at least as of now.

In the case of $D=4$ parity violating case 
\begin{equation}
L^{-2}\epsilon_{\mu_1\mu_2\mu_3\mu_4}\partial_+ X^{\mu_1}\partial_- X^{\mu_2}\partial_+^2 X^{\mu_3}
\partial_-^2 X^{\mu_4}
\end{equation}
the order $R^{-2}$ term 
$\epsilon_{\mu_1\mu_2\mu_3\mu_4}e_+^{\mu_1}e_-^{\mu_2}\partial_+^2 Y^{\mu_3}\partial_-^2 Y^{\mu_4}$, 
which we had said could be reduced to an irrelevant term in \cite{orig}, can actually
be eliminated by partial integration as it reduces to
\begin{equation}
\epsilon_{\mu_1\mu_2\mu_3\mu_4}e_+^{\mu_1}e_-^{\mu_2}\partial_{+-}Y^{\mu_3}\partial_{+-}Y^{\mu_4}
\end{equation}
due to the complete antisymmetry of the $\epsilon_{\mu_1\mu_2\mu_3\mu_4}$. 

This makes no difference for the way we have handled the irrelevant
terms. For Drummond this would make a difference as what was thought
to be an irrelevant term, necessitating a field redefinition and the
attendant induced transformation law, is actually shown to vanish.

For the rest we cannot go beyond what we already presented in
\cite{orig}. Thus in the parity violating sector there are indeed
$R^{-4},R^{-5}$ order terms.

We return later to the issue of how the irrelevant terms at order
$R^{-3}$ should be handled in the $D=4$ context also.

\subsection{Irrelevance of the non-existence of $R^{-4},R^{-5}$ terms to our original proof}
\label{irrrm4rm5}
Though we concede that we were in error in claiming that Drummond had
missed these terms, and we also concede that the absence of these
terms is very important for the issue of even higher order corrections
to the spectrum, we point out that it is irrelevant for the main
result of our paper which was to show that there are no order $R^{-3}$
corrections to the spectrum, as compared to Nambu-Goto (NG) theory.

Of course Drummond also has this result in \cite{drum}, but the
important point is he did not prove the absence of $R^{-3}$ terms. In
that paper he merely asserted, though correctly, that there were no
further action terms up to order $R^{-6}$.  Considering the importance
of the result in question, a systematic proof is absolutely
necessary. In retrospect, this is how we should have worded our
criticism of \cite{drum}.

In his response \cite{drumresp}, Drummond does prove systematically
all his earlier claims; but to do so he employs the very methods of
systematic analysis that we had developed in \cite{orig}.  Of course,
he has added some improvements here by analysing in terms of pairs of
contracted $X$ fields, and by his observation that for {\em odd} $N$
the condition $2M+2 \ge 3N$ can be improved to $2M+2\ge 3N+1$.

It is also worth pointing out that though his demonstration of the
absence of $R^{-4},R^{-5}$ order terms is certainly important, it is
not of much use until the PS transformation law is appropriately
modified as it closes only up to $R^{-4}$ order.  In general, giving
possible terms in the action without discussing the transformation
laws that would leave them invariant is incomplete. It could well be
that there are no transformation laws that leave some or all of them
invariant.

Finally, one may note that just as the absence of additional terms in
the action of order $R^{-3}$ does not automatically imply (as is
evident from \cite{drum,orig}) the absence of $R^{-3}$ corrections to
the NG spectrum, absence of $R^{-4},R^{-5}$ terms in the action also
does not translate immediately into any statement about the even
higher order corrections to the spectrum.

\section{Our alleged use of field redefinitions}
\label{afr}
Drummond has argued that, notwithstanding our statements in
\cite{orig}, we have actually used field redefinitions of the same
type used by him. Our reply to this follows (we further substantiate
comments made here in section \ref{rmf}).

Drummond cites our treatment of the $D=3$ parity violating case (and
possibly our $D=4$ parity violating case also) to make this point. If
he insists on taking the parity violating cases seriously, then his
claim that the next corrections occur only at $R^{-6}$ is already
false, as elaborated above.

Even taking the parity violating cases seriously, what we have done
does not amount to a \emph{field redefinition} but only to a
\emph{choice of field definition}. This is an important distinction
which we now explain fully.

\subsection{Choice of field definition}
\label{cfd}
The PS procedure can be stated algorithmically as follows: Firstly,
write down all possible $(1,1)$ terms (in the sense used by PS);
Secondly, discard all terms proportional to the leading order
constraints and their derivatives; Finally, use integration by parts
to relate equivalent terms.

At this point one will have terms with and without `mixed
derivatives', terms sporting mixed derivatives being what we have
called irrelevant.  The PS prescription then is to discard all
irrelevant terms and \emph{find transformation laws} that leave the
relevant terms in the action invariant.

This is still ambiguous as not all the relevant terms are independent
and some can be related to others through integration by parts and
additional irrelevant terms. The unambiguous method is to first
express all the relevant terms in terms of a particular choice of a
minimal set (this choice in itself being arbitrary) and
additional irrelevant terms. Next, one chooses a subset of the
irrelevant terms and simply drops all the rest. Finally, one finds
transformation laws that leave this combination of relevant and
irrelevant terms invariant (modulo issues of quantum equivalence
discussed earlier).

Now a particular choice of mixed derivative terms amounts to a
\emph{choice of field definition}. A different choice of the
irrelevant terms amounts to yet another choice of field
parametrisation. As long as the conditions for equivalence under field
redefinitions hold, one can go from one parametrisation to another
with the help of a \emph{field redefinition}. The transformation laws
in the new definition can be worked out as induced by the field
redefinition (this is somewhat more than mere substitution).

Let us illustrate this with an example. Take the partial set of terms
at order $R^{-2}$ to be
\begin{eqnarray}
\label{sample}
&&\alpha \frac {\partial_+^2 X\cdot\partial_-^2 X}{L}+\beta \frac {\partial_+^2 X\cdot\partial_- X
\partial_-^2 X\cdot\partial_+ X}{L^2}\nonumber\\
&+&\delta \frac {\partial_{+-}X\cdot\partial_{+-}X}{L}
+\eta \frac{\partial_{+-}X\cdot\partial_+ X
\partial_{+-}X\cdot\partial_-X}{L^2}
\end{eqnarray}
In \cite{drum} it is claimed that the two relevant actions here are
equivalent modulo total derivatives. This is not true and this has
some bearing on the issues discussed; using the identity
\begin{eqnarray}
& &\frac{\partial_+^2 X\cdot\partial_-^2 X}{L} = 
\frac {\partial_+^2 X\cdot\partial_- X
\partial_-^2 X\cdot\partial_+ X}{L^2}\nonumber\\
&+&\frac {\partial_{+-}X\cdot\partial_{+-}X}{L}
- \frac{\partial_{+-}X\cdot\partial_+ X
\partial_{+-}X\cdot\partial_-X}{L^2}\nonumber\\
&+&\partial_-\big(\frac{\partial_+^2X\cdot\partial_-X}{L}\big)
-\partial_+\big(\frac{\partial_{+-}X\cdot\partial_-X}{L}\big)
\end{eqnarray}
\eqref{sample} can be rewritten as
\begin{eqnarray}
&&\beta^\prime \frac {\partial_+^2 X\cdot\partial_- X
\partial_-^2 X\cdot\partial_+ X}{L^2} \\
&+&\delta^\prime \frac {\partial_{+-}X\cdot\partial_{+-}X}{L}
+\eta^\prime \frac{\partial_{+-}X\cdot\partial_+ X
\partial_{+-}X\cdot\partial_-X}{L^2} \nn
\end{eqnarray}
with $\beta^\prime = \beta + \alpha$,$\delta^\prime = \delta + \alpha
$ and $\eta^\prime = \eta - \alpha$. Now the choice $\delta^\prime
=0,\eta^\prime = 0$ yields one field definition, say, $X^\prime$ with
the effective action
\begin{equation}
\beta^\prime \frac {\partial_+^2 X^\prime\cdot\partial_- X^\prime
\partial_-^2 X^\prime\cdot\partial_+ X^\prime}{{L^\prime}^2}
\end{equation}
while the choice $\beta^\prime = \alpha$,$\delta^\prime = \alpha$ and
$\eta^\prime = -\alpha$ gives another field definition, say, $X^*$
with the effective action
\begin{equation}
\alpha \frac{\partial_+^2 X^*\cdot\partial_-^2 X^*}{L^*}  
\end{equation}
Thus specific choices for coefficients of the mixed derivative terms
merely pick out specific parametrisations and have nothing to do with
redefinitions. Having chosen a particular parametrisation, one has to
work out the transformation laws leaving the action invariant, the
consequent stress tensors, and so on.

In this particular example, though the two forms of relevant terms are
related by a non-trivial field redefinition, the transformation laws
for the two cases are identical. In fact there are families of field
redefinitions that do not induce any additional terms in the
transformation laws; for examples see section \ref{redefnt}. Of
course, in generic cases field redefinitions induce additional terms
in the transformation laws.

Care has to be taken to carry through this procedure consistently to
all orders. This has to do with the fact that with field redefinitions
of order $\epsilon$ there is a residual error of order $\epsilon^2$
which should be included in the analysis of the most general effective
action to be carried out to the next higher order.

\subsection{Other remarks regarding field redefinitions}
\label{morefred}
In \cite{orig} we had remarked that ``subteleties regarding the field
redefinition in eqns \cite{drum}(2.17-2.19) and how this leads to a
non-trivial energy momentum tensor are ignored.'' To this Drummond's
response \cite{drumresp} has been: ``Their claim that the field
redefinition obscures the derivation of the energy momentum tensor in
\cite{drum} is spurious - one simply rewrites expression (2.16) of
\cite{drum} in terms of ${\tilde Y}$.''  Furthermore, Drummond claims
that questions of changes in the path integral measure are unnecessary
for the field redefinitions made by him in \cite{drum}.

As for issues of measures and jacobians, we have tried to address
these in section \ref{fredefj} above. It appears that indeed these may
not be issues of consequence to the orders required in proving the
absence of corrections to the spectra in \cite{drum} and \cite{orig},
but it is clear that these are matters that could not be merely
asserted.

Turning now to the above rejection of our claim as spurious, there are
some subtle issues here which we now discuss.  We feel that Drummond
has not quite appreciated our point about ``\dots how this leads to a
nontrivial energy momentum tensor \dots''.  It may be that our wording
was opaque. It is of course true that substitution of ${\tilde Y}$ in
the PS energy momentum tensor yields his eqn.(2.20), but the subtlety
we were referring to had to do with the fact that his redefinition
eqn.(2.19) reduced the PS action modulo the terms of order higher than
$R^{-3}$ to the particular \emph{free action}
\begin{equation}
\label{simplefree}
{\cal L}^{(0)} = \frac{1}{4\pi a^2} \partial_+{\tilde Y}\cdot\partial_-{\tilde Y}
\end{equation}
and looking at this action, the casual reader would then have expected
the standard stress tensor
\begin{equation}
\label{trivial}
-\frac{1}{2a^2}\partial_-{\tilde Y}\cdot\partial_-{\tilde Y}
.\end{equation} 
Instead he has a different and non-trivial energy momentum
tensor. What we meant was that he has ignored the issue of why
\er{trivial} is not the correct choice and why the non-trivial energy
momentum tensor he obtains by substituting his field redefinition is
\emph{the} correct choice. Clearly the issue we had in mind is not so
spurious.  We in fact resolve this issue in section \ref{rmf}.

It is indeed not enough, as Drummond says, for the theory to remain
conformal with critical central charge and non-trivial energy momentum
tensor. Under field redefinitions central charge certainly cannot
change, and neither can the conformal nature of a theory; this is
somewhat beside the point, as the issue is really whether or not there
are nontrivial effects on the spectrum.

An additional claim made by Drummond is that our procedure of
iteratively solving field equations is exactly equivalent to his
procedure involving field redefinition.
Drummond does make some astute and interesting remarks about the
similarities between the two methods, but we do not agree with 
his assertion that they are equivalent for the following reasons.

In our approach, based on the iterative solution to the EOM, we
have no need to rework either the action or the transformation laws.
In Drummond's approach, both of these have to be done, though he did
not compute the induced transformation law (see section
\ref{redefnt}); he needs to compute the action to know about the
two-point function relevant for the OPE calculation. He also needs to
compute the transformation law to know whether the energy-momentum
tensor obtained by substitution is the canonical one or not.

In our method, as we explain in section \ref{cfd}, we make use of no
field redefinition. In consequence, there are no issues of quantum
equivalence. In this respect, differences between the two procedures
would be more pronounced at higher order.  Indeed, iterative solutions
to field equations are possible (in principle) to any order, but as
Drummond himself admits, his trick really works only at order
$R^{-3}$.

In general solving the full EOM iteratively involves non-locality.  It
is a fortuitous circumstance here (perhaps because of 2-d) that it is
quasi-local. Field redefinitions are typically local (not necessarily
point transformations) by contrast, with nonlocal field redefinitions
being a largely uninvestigated and difficult subject.

All one could have concluded had one designed a field redefinition
based on the iterative solution to the full equations of motion is
that ${\tilde Y}$ would be solution of the EOM of \er{simplefree}, but
of course that does not guarantee that the action is given by
\er{simplefree}. There are many actions (including the quadratic part
of the terms in PS) whose EOM is satisfied by such a redefined
field. Significantly, they generically have different two-point
functions and consequently different OPEs. It also does not guarantee
that the transformation law is the standard one associated with free
actions.  One would have had to recompute both these, though Drummond
did not do the latter. In his case, while the induced action turns out
to be free, the induced transformation law turns out to be
non-trivial.  Our method did not require the consideration of these
issues.

\subsection{Further points on field redefinitions}
\label{redefnt}
All field redefinitions are \emph{ambiguous} up to terms of the type
\begin{equation}
\Delta X^\mu = N^\mu, \qquad N\cdot E = 0
\end{equation}
where $E^\mu$ is the EOM.

Not all field redefinitions induce changes in transformation laws. 
Some examples are:
\begin{eqnarray}
\Delta_1 X^\mu &=& \frac{\partial_{+-}X^\mu}{L} \nn \\
\Delta_2 X^\mu &=& \frac{(\partial_+X^\mu\partial_{+-}X\cdot\partial_-X+\partial_-X^\mu
\partial_{+-}X\cdot\partial_+X}{L^2} \nn\\
\Delta_3 X^\mu &=& \frac{(\partial_+X^\mu\partial_{+-}X\cdot\partial_-X-\partial_-X^\mu
\partial_{+-}X\cdot\partial_+X}{L^2} \nn
\end{eqnarray}
It should be noted that $\Delta_3 X^\mu$ is of the form of an
ambiguity. It is so even when $L^{-2}$ is replaced by any power of
$L$, but the transformation law is unaffected only for $L^{-2}$.
Only the choice $L^{-2}$ maintains the $(0,0)$ character of $X^\mu$.

\section{OPE of stress tensor and Virasoro algebra to higher order}
\label{ope}
Drummond criticises by saying ``It should be pointed out that DM do
not bother to verify the operator product of the energy-momentum
tensor and hence that the expression for the Virasoro generators
really satisfies the Virasoro algebra to the next order.''  In fact,
slightly more can be said; although we did not mention it, we had
noticed that the variation of the \emph{entire} PS action, eqn (1) of
\cite{orig}, without truncating to any order, under the \emph{entire}
PS variation, eqn(2) of \cite{orig}, again without any truncations,
has only $\beta^2$ terms and these are of order $R^{-4}$.
Furthermore, the untruncated PS transformation has the closure
property of Virasoro algebra also to order $R^{-4}$ as is very easily
verified. The terms that spoil closure are again $\beta^2$ terms.
Taken together these guarantee that there are no issues with either
the OPE of stress tensors or of the validity of the Virasoro algebra
generated by their moments to the order to which we extended the PS
results.  The fact that our higher order energy-momentum tensor is
conserved is another consistency check.

\section{Our dropping order $R^{-4}$ terms}
\label{rmf}
Drummond has raised objections to our dropping order $R^{-4}$ terms in
investigating the possible $R^{-3}$ corrections to the spectrum. His
main argument here is that in his field definition he has shown that
the PS action is of order $R^{-4}$ and that neglecting these $R^{-4}$
terms can only be done after taking due account of the changes in the
transformation law induced by the field redefinition that allowed
reducing the PS action to $O(R^{-4})$. Consequently, he claims, our
treatment of the parity violating terms where we simply drop the
$R^{-2},R^{-3}$ terms as being reducible to mixed derivative terms is
incorrect.

Again, as stated earlier, what we have done is drop all the irrelevant
terms, and find the symmetry variations for what is retained. This
means that, to the order we were originally interested in, the only
relevant action terms are the PS term and the free action; the PS
transformation law leaves them invariant, actually to order
$R^{-3}$. Noting this, nothing more need be done.

As we have explained, our treatment merely amounts to a specific
choice of field parametrisation and not to a redefinition.  We feel
that the distinction we made between these in section \ref{cfd} is an
important one.  In contrast to this straightforward procedure,
Drummond suggests that we should have first determined the
transformation laws that would leave invariant the free and PS terms
along with the irrelevant terms in the parity violation action, then
carried out the field redefinition that would remove the irrelevant
terms, and finally computed the modifications to the transformation
laws.  We now show through explicit calculation of what he is
proposing that these extra ingredients are totally unnecessary and our
treatment of these and other similar terms in \cite{orig} is perfectly
legitimate.

For this purpose consider a generic irrelevant term
\begin{equation}
{\cal L}_{\textup{irr}} = {\cal F}^\mu \partial_{+-}X_\mu
\end{equation}
where ${\cal F}^\mu$ can be any general expression constructed out of
derivatives of $X^\mu$. In particular it can also contain additional
mixed derivative terms.  Now we wish to modify the transformation law
so that it leaves ${\cal L}_{\textup{free}}+{\cal L}_{\textup{irr}}$
invariant.  That this can always be done follows from the fact that
the variation of the EOM is proportional to the EOM (in a functional
sense) which is just another way of saying the EOM is covariant under
symmetry variations.  Now we evaluate the variation of ${\cal
L}_{\textup{irr}}$ under
\begin{equation}
\delta^{(0)}X^\mu = \epsilon^-\partial_- X^\mu
,\end{equation}
yielding
\begin{eqnarray}
\delta^{(0)}{\cal L}_{\textup{irr}} &=& \delta^{(0)}{\cal F}^\mu\partial_{+-}X_\mu
      +{\cal F}^\mu\delta^{(0)}(\partial_{+-}X_\mu)\nonumber\\
&=& 
\delta^{(0)}{\cal F}^\mu\partial_{+-}X_\mu-\epsilon^-\partial_-{\cal F}^\mu\partial_{+-}X_\mu)
,\end{eqnarray}
where we have dropped total derivative terms as usual. Therefore the
modification to the transformation law under which 
${\cal L}_{\textup{free}}+{\cal L}_{\textup{irr}}$ will be 
invariant is
\begin{equation}
\delta^\prime X^\mu = 2\pi a^2(\delta^{(0)}{\cal F}^\mu-\epsilon^-\partial_-{\cal F}^\mu)
.\end{equation}
On the other hand the field redefinition needed to remove the irrelevant term is
\begin{equation}
\Delta X^\mu = -2\pi a^2 {\cal F}^\mu
.\end{equation}
This induces the additional terms
\begin{eqnarray}
{\tilde\delta}X^\mu&=&\delta^{(0)}\Delta X^\mu - \epsilon^-\partial_-(\Delta X^\mu)\nonumber\\
&=& 
- 2\pi a^2(\delta^{(0)}{\cal F}^\mu-\epsilon^-\partial_-{\cal F}^\mu)
\end{eqnarray}
and this contribution exactly cancels the modification demanded by the irrelevant term.

Generalisation of these arguments to include an arbitrary number of
relevant terms in the action is straightforward as long as one keeps
track of the order of terms carefully.

We conclude that what Drummond claims to be the correct procedure is
completely equivalent to simply dropping all irrelevant terms and
working with the transformation laws that leave only the relevant
terms invariant, which is exactly what we did in \cite{orig}.
Thereafter dropping the irrelevant parity violating terms one was left
with just the free action and PS terms as relevant and as the PS
transformation laws expanded to the next higher order left them
invariant, there was nothing wrong with our methodology.

In support of his argument Drummond says that if he had
followed a similar procedure he would have dropped the entire PS
action which, according to him, is not permissible.
In the following section we instead take this suggestion seriously.

\section{Dropping the PS action}
In the previous section we responded to some criticisms made
by Drummond about our dropping of $\Pcm{O}(R^{-4})$ terms.
He suggested that our methods could lead to our discarding
of the whole PS action.
In particular his argument seems to be that the PS action
has left its trace in terms of a non-trivial energy momentum tensor
and a non-trivial transformation law (though he did not work this out)
remembering the parameters (i.e. $\beta$) of the irrelevant terms.

Following this argument, according to him our irrelevant terms also
could in principle have left their nontrivial traces, which are simply
not there in our way of handling things.  Consider, though, that we
have explicitly and quite generally proved in the previous section
that the field redefinitions exactly compensate any modifications to
the transformation laws brought forth by the irrelevant terms: How can
these two apparently contradicting situations be reconciled?

Before we go on we wish to categorically state that the
$R^{-4}$-order terms are totally irrelevant for our analysis in
\cite{orig}. This is also true for Drummond's analysis in
\cite{drum}. The simple reason for this is that the variation of such
terms under the PS transformation law can \emph{at most} be of order
$R^{-3}$ and consequently their contribution to the Virasoro
generators can also be at most $R^{-3}$.
The imprints in the transformation law that Drummond is
possibly referring to are all from $R^{-2}$ or $R^{-3}$ terms.

Now, let us go on and examine the case of PS action more carefully.  If
we had simply dropped the irrelevant terms from the PS action we would
have ended up only with (modulo $R^{-4}$ and higher order terms)
\begin{equation}
\frac{1}{2\pi a^2}\int \partial_+Y\cdot\partial_- Y
\end{equation}
Now according to our prescription we should just work with this
relevant action and the transformation laws leaving it invariant.
Somewhat surprisingly, it turns out that not just the standard
\begin {equation}
\delta_-^{(0)} Y^\mu = \epsilon^- \partial_-Y^\mu
.\end{equation}
but {\em at least} the entire two-parameter family
\begin{eqnarray}
\label{biggertrans}
{\bar\delta}^{(0)}_- Y^\mu &=& \epsilon^- \partial_- Y^\mu
 + \epsilon^- R~e_-^\mu \\
&+&\frac{\beta^\prime a^2}{R} \partial_-^2\epsilon^-~e_+^\mu
+\frac{2\beta^\prime a^2}{R^2}\partial_-^2\epsilon^- e_+^\mu e_+\cdot\partial_- Y \nn
\end{eqnarray}
leaves the above action invariant (exactly). Also, they constitute the
conformal group (for fixed $R$ and $\beta'$) as can be verified by
\begin{equation}
[\delta_-(\epsilon^-_1),\delta_-(\epsilon^-_2)] =  \delta_-(\epsilon^-_{12});~~~\epsilon^-_{12} = \epsilon^-_1\partial_-\epsilon^-_2 - \epsilon^-_2\partial_-\epsilon^-_1
\end{equation}
It should be emphasised that at this stage the parameter
$\beta^\prime$ has nothing to do with $\beta$.

The resulting canonical energy-momentum tensor is indeed the same with
$\beta$ replaced by $\beta^\prime$ as eqn(19) of \cite{orig} or
eqn(2.20) of \cite{drum} (after correcting a sign error). The central
charge depends on the free parameter $\beta^\prime$ and can be
adjusted for consistency in all dimensions.

On the other hand, if we had included the irrelevant terms of
Drummond, the transformation law that would have left the above
combination of relevant and irrelevant terms invariant would have been
modified to
\begin{equation} 
\label{total}
\delta_-^{tot} = {\bar\delta}_-^{(0)}Y^\mu+
\frac{\beta a^2}{R} \partial_-^2\epsilon^-~e_+^\mu
+\frac{2\beta a^2}{R^2}\partial_-^2\epsilon^- e_+^\mu e_+\cdot\partial_- Y
\end{equation}

If now one had done a field redefinition to remove the irrelevant
terms there would be additional terms induced in the above
equation. As shown above their effect would be to cancel the
$\beta$-dependent terms and just leave \er{biggertrans}.

If Drummond had computed the transformation law to which the
PS-transformation law would have been modified by the field
redefinition (which he actually did not, for all his admonitions to us
about the importance of working out the induced terms in
transformation laws from field redefinitions) he would have found them
to be just \er{biggertrans} with $\beta^\prime$ replaced by $\beta$,
modulo some harmless terms arising out of his not having paid
attention to the many ambiguities in field redefinitions.

The question now is: How did $\beta^\prime$ get to be replaced by
$\beta$ which is a parameter belonging entirely to the irrelevant
terms?  The resolution is the following. According to the above
construction what would leave the combination of relevant and
irrelevant terms of this example would be \er{total}, but this is
clearly not the PS-transformation law expanded to suitable orders!
Nevertheless in Drummond's scheme of things he would have taken this
to be PS-transformation suitably expanded. This is consistent with
\er{total} \emph{if and only if $\beta^\prime$ had been equated to $\beta$}.

It seems then that, contrary to folklore, a free bosonic string theory
with an adjustable central charge can be made consistent provided the
conformal transformation law is chosen appropriately.  This can be
viewed as an alternative approach to the spectrum of free strings that
appears consistent in all dimensions.  How far such an approach to
string theory can be extended further is currently under investigation
\cite{freebos}.

\section{Conclusion}
\label{concl}
We have examined a variety of issues bearing on our earlier work
proving the absence of $R^{-3}$ corrections to the Nambu-Goto spectrum
in effective string theories in light of Drummond's criticisms of
our paper. Here we summarise the main points.

We agree that as far as classical equivalence of field theories is
concerned, terms alleged to have existed at order $R^{-4},R^{-5}$ in
effective string theories, can really be transformed away as
irrelevant terms by integration by parts and field redefinitions in
the parity conserving sector.  In the parity violating sector no such
conclusions can be drawn even classically. Quantum-mechanically,
one has to exercise greater care in the transformation of measure
(or equivalent technique in the canonical formalism) and at this point
it is not at all clear whether order-$R^{-4},R^{-5}$ terms play a r\^ole. 
In his original work Drummond only asserted the absence of these terms.
While he did prove them systematically in his response to our work,
these systematic proofs made use of the very techniques we developed 
in our original paper.
The existence or otherwise of these terms is irrelevant for
our original proof of the absence of $R^{-3}$ corrections to
NG-spectrum in effective string theories.
Drummond also had this result but he only asserted and did
not prove the crucial absence of additional terms in the action at
$R^{-3}$ order (apart from what are already contained in the PS action).
We stress that what we really showed in \cite{orig} was that given the
PS action and transformation laws there are no $R^{-3}$ corrections to
the NG-spectrum staying entirely within the PS parametrisation.

We have carefully analysed Drummond's remarks about field redefinition
and show them to be incorrect through general proofs as well as
explicit constructions.

In conclusion we wish to state that at least as far as we are
concerned this debate with Drummond has enhanced our understanding of
a variety of important issues connected with effective string
theories.

\end{document}